\newfam\msbfam
\font\twlmsb=msbm10 at 12pt
\font\eightmsb=msbm10 at 8pt
\font\sixmsb=msbm10 at 6pt
\textfont\msbfam=\twlmsb
\scriptfont\msbfam=\eightmsb
\scriptscriptfont\msbfam=\sixmsb

\centerline{\bf EINSTEIN-CARTAN THEORY AND GAUGE SYMMETRY} 

\

\centerline{M. Socolovsky*}

\

\centerline{\it  Instituto de Ciencias Nucleares, Universidad Nacional Aut\'onoma de M\'exico}
\centerline{\it Circuito Exterior, Ciudad Universitaria, 04510, M\'exico D. F., M\'exico} 

\

{\bf Abstract} {\it We argue that the non gauge invariant coupling between torsion and the Maxwell or Yang-Mills fields in Einstein-Cartan theory can not be ignored. Arguments based in the existence of normal frames in neighbourhoods, and an approximation to a $\delta$-function, lead to gauge invariant observables.}

\

As is well known, it is usually claimed that the electromagnetic and the Yang-Mills fields do not couple to torsion in the Einstein-Cartan theory, due to the fact that the corresponding local gauge symmetries $U(1)$ or $SU(2)$ (or $SU(3)$, etc.) are violated at the level of the Cartan equation of motion (Hehl {\it et al}, 1976). For the Maxwell field (for simplicity we shall restrict to the abelian case) the Cartan equation is $$T^\mu_{\rho\sigma}+\delta^\mu_\rho T_\sigma-\delta^\mu_\sigma T_\rho=-{{l}\over{2}}S^\mu_{\rho\sigma} \eqno{(1)}$$ where $$T^\mu_{\rho\sigma}={{1}\over{2}}(\Gamma^\mu_{\rho\sigma}-\Gamma^\mu_{\sigma\rho}) \eqno{(2)}$$ is the torsion tensor, $T_\rho=T^\mu_{\rho\mu}$, $l={{G}\over{c^4}}$ and $$S^\mu_{\rho\sigma}={F^\mu}_\rho A_\sigma-{F^\mu}_\sigma A_\rho \eqno{(3)}$$ is the canonical spin density tensor of the electromagnetic field, obtained from Noether theorem applied to the Einstein-Maxwell Lagrangian density $${\cal L}_{E-M}=-{{1}\over{4}}F^{\mu\nu}F_{\mu\nu} \eqno{(4)}$$ with the electromagnetic field strength given by the {\it curvature} of the abelian connection $A=A_\mu dx^\mu$, $$F=dA={{1}\over{2}}F_{\mu\nu}dx^\mu\wedge dx^\nu\eqno{(5)}$$ with $$F_{\mu\nu}=\partial_\mu A_\nu-\partial_\nu A_\mu=(\partial_{[\mu}A_a-\omega^b_{\mu a}A_b){e_{\mu]}}^a+A_bT^b_{\mu\nu},\eqno{(6)}$$ where $e^a={e_\mu}^a dx^\mu$ and ${\omega^b}_a=\omega^b_{\mu a}dx^\mu$ are respectively the tetrad and spin connection 1-forms, $A^a={e_a}^\nu A_\nu$ with ${e_a}^\nu{e_\nu}^b=\delta^b_a$, and $[\dots]$ denotes antisymmetrization in the $\mu\nu$ indices. Clearly, $F$ is gauge invariant, since under $$A\to A^\prime=A+d\lambda,  \ \ F\to F^\prime=F+d^2\lambda=F.\eqno{(7)}$$ $F_{\mu\nu}$ in the r.h.s. of (6) results from the {\it minimal coupling} between the electromagnetic connection (considered here as a matter field, though massless) and the spacetime connection: $$dA_a\to DA_a=dA_a-{\omega^b}_aA_b,\eqno{(8)}$$ (Benn, Dereli, and Tucker, 1980), and the expression for torsion $$T^a=de^a+{\omega^a}_b\wedge e^b={{1}\over{2}}T^a_{\mu\nu}dx^\mu\wedge dx^\nu, \ \ T^a_{\mu\nu}={e_\rho}^aT^\rho_{\mu\nu}.\eqno{(9)}$$

\

The total Einstein-Maxwell action including the pure gravity term is $$S=S_G+S_{E-M}=l\int d^4x \ e(R+{\cal L}_{E-M}), \eqno{(10)}$$ where $e=\sqrt{-det(g_{\mu\nu})}$ and $R=\eta^{bc}R^a_{b\mu\nu}{e_a}^\mu {e_c}^\nu$ is the Ricci scalar. (Greek indices are raised and lowered with the pseudo-Riemannian metric $g_{\mu\nu}$, while latin indices are rased and lowered with the Minkowski metric $\eta_{ab}=diag(1,-1,-1,-1)$.) Equation (1) is obtained from (10) by variation with respect to the spin connection. (For details see e.g. Socolovsky, 2012.) 

\

The relation between the spin connection, tetrads, and Christoffel symbols is $$\omega^b_{\mu a}={e_\rho}^b\partial_\mu{e_a}^\rho+{e_\rho}^b{e_a}^\nu\Gamma^\rho_{\mu\nu},\eqno{(11)}$$ where $$\Gamma^\rho_{\mu\nu}=(\Gamma_{LC})^\rho_{\mu\nu}+K^\rho_{\mu\nu}, \eqno{(12)}$$ $\Gamma_{LC}$ being the Levi-Civita connection of general relativity (GR) (Einstein, 1956), and $K^\rho_{\mu\nu}$ the {\it contortion tensor} given by $$K^\rho_{\mu\nu}=(K_A)^\rho_{\mu\nu}+(K_S)^\rho_{\mu\nu}, \eqno{(13)}$$ $$(K_A)^\rho_{\mu\nu}=T^\rho_{\mu\nu}, \eqno{(14)}$$ $$(K_S)^\alpha_{\mu\nu}=g^{\alpha\rho}(T^\lambda_{\rho\mu}g_{\lambda\nu}+T^\lambda_{\rho\nu}g_{\lambda\mu}). \eqno{(15)}$$ By the gauge transformation (7), the gauge variation of $S^\mu_{\nu\rho}$ is $$\delta_{g.tr.}(S^\mu_{\rho\sigma})=2{F^\mu}_{[\rho}\partial_{\sigma]}\Lambda. \eqno{(16)}$$ So, the spin density tensor of the electromagnetic field and therefore the Cartan equation (1) are not gauge invariant.

\

It is important to remark here that in the flat Minkowski spacetime of special relativity in the context of classical and quantum electrodynamics, $S^\mu_{\nu\rho}$ is also given by (3). That is, already in special relativistic field theory the spin density tensor of the electromagnetic field is not gauge invariant (Bogoliubov and Shirkov, 1980) and, in contradistinction with the energy-momentum density tensor, there is no recipe similar to the Rosenfeld-Belinfante prescription (Landau and Lifshitz, 1975) for this case, to make the spin density gauge invariant. The reason of why even starting from a gauge and Lorentz invariant Lagrangian $F_{\mu\nu}F^{\mu\nu}$, the Noether theorem leads to a non gauge invariant spin density tensor, is that the global Lorentz and local gauge transformations of the gauge potential do not commute. In fact, let $\lambda(x)$ be the local gauge transformation (g.tr.), and $\Lambda_\mu^\nu$ the global Lorentz rotation (r.); then: $$A_\mu(x)\buildrel{\lambda}\over\longrightarrow A_\mu(x)+\partial_\mu\lambda(x)\buildrel{\Lambda}\over\longrightarrow\Lambda_\mu^\nu(A_\mu(x)+\partial_\mu\lambda(x))\equiv (A_\mu^{g.tr.})^{r.},$$ while $$A_\mu(x)\buildrel{\Lambda}\over\longrightarrow\Lambda_\mu^\nu A_\nu(x)\buildrel{\lambda}\over\longrightarrow\Lambda_\mu^\nu A_\nu(x)+\partial_\mu\lambda(x)\equiv(A_\mu^{r.})^{g.tr.}.$$ Therefore $$((A^{g.tr.})^{r.}-(A^{r.})^{g.tr.})_\mu=(\Lambda_\mu^\nu\partial_\nu-\partial_\mu)\lambda(x).$$ For an infinitesimal Lorentz rotation $\Lambda_\mu^\nu=\delta_\mu^\nu+\epsilon_\mu^\nu$ and so $$[r.,g.tr.](A_\mu)=\epsilon_\mu^\nu\partial_\nu\lambda.\eqno{(17)}$$ However, observable quantities like left and right circular polarized light (helicity in the quantum theory), turn out to be gauge invariant.

\

The usual solution to the conflict between E-C torsion and gauge invariance, is to declare that torsion and the Maxwell (or Yang-Mills) field do not ``see'' each other i.e. they do not couple. But this argument is extremely weak -if not wrong- by several reasons: i) The minimal coupling (8) is not a mere prescription or convention to couple matter to geometry, but it is the local version of the global definition of the covariant derivative in the context of fibre bundle theory    (see Appendix). ii) Locally, torsion appears {\it additively} in (12) through the contortion tensor, with the first term being the Levi-Civita connection; the coupling of $\Gamma_{LC}$ to the Maxwell field in GR is well tested experimentally, so that there is no reason to declare the vanishing of the coupling of $K^\rho_{\mu\nu}$ with $A_\lambda$, moreover in the absence of any coupling constant to be switched-off. iii) In quantum field theory to lowest order in perturbation theory, torsion couples to photons through the 1-loop electron-positron pair (de Sabbata, 1997), with the coupling torsion-Dirac field explicitly gauge invariant. It is quite unnatural that the $K-A$ coupling vanishes in the classical limit.

\

It is well understood that the decomposition of fields in positive and negative frequency parts it is not possible in curved spacetime unless there exists a time-like Killing vector field (Carroll, 2004). However, according to Iliev (Iliev, 1996), locally, i.e. in the neighbourhood of any point, we can choose normal frames and assume that in these frames the special relativistic equations hold, in particular the local validity of the above mentioned decomposition. Approximating an integral over a finite volume $V$ by a $\delta$-function, leads us to a gauge invariant expression for the components of the polarization vector of the Maxwell field. The procedure is similar to the case of field theory in Minkowski spacetime, and we present it for completeness.

\

The solution of the algebraic equation (1) is $$T^\mu_{\rho\sigma}={{l}\over{2}}(S^\mu_{\rho\sigma}+{{1}\over{2}}(\delta^\mu_\sigma S_\rho-\delta^\mu_\rho S_\sigma)) \eqno{(18)}$$  with $S_\rho=S^\nu_{\rho\nu}$. For $\mu=0$ and $\rho\sigma=jk$ one has $$T^0_{jk}={{l}\over{2}}(S^0_{jk}+{{1}\over{2}}(\delta^0_k S_j-\delta^0_j S_k))={{l}\over{2}}S^0_{jk} \eqno{(18)}$$ and therefore the spatial components of the polarization vector of the Maxwell field in a volume element $V$ are given by $$S_i=\epsilon_{ijk}\int_Vd^3\vec{x} \ S^0_{jk}={{4c^4}\over{G}}\int_Vd^3\vec{x} \ T^0_{jk}\eqno{(20)}$$ i.e. $$S_1=2\int_Vd^3\vec{x} \ S^0_{23}={{4c^4}\over{G}}\int_Vd^3\vec{x} \ T^0_{23}, \ S_2=2\int_Vd^3\vec{x} \ S^0_{31}={{4c^4}\over {G}}\int_Vd^3\vec{x} \ T^0_{31},$$ $$  S_3=2\int_Vd^3\vec{x} \ S^0_{12}={{4c^4}\over{G}}\int_Vd^3\vec{x} \ T^0_{12} \ . \eqno{(21)}$$ One choses the gauge in which $$\partial^\mu A_\mu=0, \ A^0=0. \eqno{(22)}$$ The second condition leads to $$S^0_{jk}=A_k\dot{A}_j-A_j\dot{A}_k \eqno{(23)}$$ with $\dot{A}={{\partial}\over{\partial x^0}}A_j$. Then $$S_i=\epsilon_{ijk}\int_Vd^3\vec{x} \ (A_k\dot{A}_j-A_j\dot{A}_k) \ . \eqno{(24)}$$ In $V$ we assume the validity of the decomposition in positive and negative frequencies $$A_k(x)=A_k^+(x)+A_k^-(x), \ k=1,2,3, \eqno{(25)}$$ $$A_j^{\pm}(x)={{1}\over{(2\pi)^{{{3}\over{2}}}}}\int{{d^3\vec{k}}\over{\sqrt{2\vert\vec{k}\vert}}} \ e^{{\pm}i(\vert \vec{k}\vert t-\vec{k}\cdot\vec{x})}A_j^{\pm}(\vec{k}). \eqno{(26)}$$ Making the approximation $$\int_V d^3\vec{x} \ e^{i\vec{l}\cdot\vec{x}}\cong \delta^3(\vec{l}) \eqno{(27)}$$ a straightforward calculation leads to $$S_l=2i\epsilon_{lmn}\int d^3\vec{k} \ A_m^+(\vec{k})A_n^-(\vec{k}) \eqno{(28)}$$ i.e. $$\vec{S}={{2c^4}\over{G}}\vec{T}^0=2i\int d^3\vec{k} \ \vec{A}^+(\vec{k})\times\vec{A}^-(\vec{k}), \eqno{(29)}$$ with $$\vec{T}^0=((\vec{T}^0)_1,(\vec{T}^0)_2,(\vec{T}^0)_3)), \ (\vec{T}^0)_i=\epsilon_{ijk}\int_V d^3\vec{x} \ T^0_{jk}. \eqno{(30)}$$ The first of the gauge conditions (22) implies the transversality condition for the Fourier components of the Maxwell potential: $$\vec{k}\cdot\vec{A}^{\pm}(\vec{k})=0. \eqno{(31)}$$ Choosing an orthonormal tetrad in momentum space, $$e^a(\vec{k})={e_\mu}^a(\vec{k})dk^\mu, \ {e_\mu}^0(\vec{k})=(1,\vec{0}), \ {e_\mu}^i(\vec{k})=(0,\hat{e}_i), \ i=1,2, \ {e_\mu}^3(\vec{k})=(0,\hat{k}), \ \hat{e}_1\times\hat{e}_2=\hat{k}, \ (e^a,e^b)=\eta^{ab}, \eqno{(32)}$$ one has the decomposition $$A_\mu^{\pm} (\vec{k})=\sum_{a=0}^3\alpha_a^{\pm}{e_\mu}^a(\vec{k})\eqno{(33)}$$ with $$(A^+(\vec{k}),A^-(\vec{k}))=\sum_{\mu,\nu=0}^3\eta^{\mu\nu}A_\mu^+(\vec{k})A_\nu^-(\vec{k})=\alpha_0^+(\vec{k})\alpha_0^-(\vec{k})
-\vec{\alpha}^+(\vec{k})\cdot\vec{\alpha}^-(\vec{k})=-\vec{\alpha}^+_T (\vec{k})\cdot\vec{\alpha}^-_T(\vec{k}) \eqno{(34)}$$ with $\vec{\alpha}^{\pm}_T(\vec{k})=(\alpha^{\pm}_1(\vec{k}),\alpha^{\pm}_2(\vec{k}))$, since by the transversality condition (31), $\alpha_0^+(\vec{k})\alpha_0^-(\vec{k})-\alpha_3^+(\vec{k})\alpha_3^-(\vec{k})=0.$  

\

In terms of the $\alpha^{\pm}_a$ components, $$S_l=\epsilon_{ljk}\int d^3\vec{k} \ {e_j}^b(\vec{k})\alpha^+_b(\vec{k}){e_k}^c(\vec{k})\alpha^-_c(\vec{k})=2i\int d^3\vec{k} \ ((\hat{e}^b\alpha^+_b(\vec{k}))\times (\hat{e}^c(\vec{k})\alpha^-_c(\vec{k})))_l$$ i.e. $$\vec{S}=2i\int d^3\vec{k} \ \vec{\alpha}^+(\vec{k})\times \vec{\alpha}^-(\vec{k}). \eqno{(35)}$$ In particular, $$S_3=2i\int d^3\vec{k} \ (\alpha^+_1(\vec{k})\alpha^-_2(\vec{k})-\alpha^+_2(\vec{k})\alpha^-_1(\vec{k})). \eqno{(36)}$$ Making the Bogoliubov transformation: $$\alpha^{\pm}_1(\vec{k})={{1}\over{\sqrt{2}}}(\beta^{\pm}_1(\vec{k})+\beta^{\pm}_2(\vec{k})), \ \alpha^{\pm}_2(\vec{k})={\pm}{{i}\over{\sqrt{2}}}(\beta^{\pm}_1(\vec{k})-\beta^{\pm}_2(\vec{k})), \ \alpha_3^{\pm}(\vec{k})=\beta_3^{\pm}(\vec{k}), \eqno{(37)}$$ one obtains $$S_3=2\int d^3\vec{k} \ (\beta_1^+(\vec{k})\beta_1^-(\vec{k})-\beta_2^+(\vec{k})\beta_2^-(\vec{k})), \eqno{(38)}$$ which is the standard gauge invariant decomposition of the component of the polarization vector of the electromagnetic field along the direction of motion of the wave (+ and - helicity states after quantization). 

\

{\bf Acknowledgments}

\

The author thanks the Instituto de Ciencias de la Universidad Nacional de General Sarmiento, Pcia. de Buenos Aires, Argentina, and the Instituto de Astronom\'\i a y F\'\i sica del Espacio, Universidad de Buenos Aires y CONICET, Argentina, for their hospitality during the sabbatical period. Also, to Dr. Rafael Ferraro for enlightening discussions on the subject. This work was partially supported by the project PAPIIT IN101711-2, DGAPA-UNAM, M\'exico.

\

{\bf References}

\

Benn, I. M., Dereli, T., and Tucker, R. W. (1980). Gauge field interactions in spaces with arbitrary torsions, {\it Physics Letters} {\bf 96B}, 100-104.

\

Bogoliubov, N.N. and Shirkov, D.V. (1980). {\it Introduction to the Theory of Quantized Fields}, Wiley, pp.50-51.

\

Carroll, S. (2004). {\it Spacetime and Geometry. An Introduction to General Relativity}, Addison Wesley, p. 396.

\

Einstein, A. (1956). {\it The Meaning of Relativity}, Chapman and Hell.

\

Hehl, F. W., von der Heyde, P., Kerlick, G. D., and Nester, J. M. (1976). General relativity with spin and torsion: Foundations and Prospects, {\it Reviews of Modern Physics} {\bf 48}, 393-416.

\

Iliev, B.Z.(1996). Normal Frames and the Validity of the Equivalence Principle: II. Cases of a Neighbourhood and at a Point, {\it Journal of Physics A: Math.Gen.} {\bf 29}, 6895-6901; arXiv: gr-qc/9608019.

\

Landau, L. D., and Lifshitz, E. M. (1975). {\it The Classical Theory of Fields, Course of Theoretical Physics, Vol. 2}, Elsevier, pp. 83-84.

\

de Sabbata, V. (1997). Evidence for torsion in gravity?, in {\it Spin Gravity: Is it possible to give an experimental basis to torsion?}, International School of Cosmology and Gravitation, XV Course, Erice, Italy, eds. P. G. Bergmann {\it et al}, World Scientific, pp. 52-85.

\

Socolovsky, M. (2012). Fibre Bundles, Connections, General Relativity, and the Einstein-Cartan Theory, {\it Advances in Applied Clifford Algebras}, {\bf 22}, 837-872 (Part I), 873-909 (Part II); arXiv: gr-qc 1110.1018v1.

\

{\bf Appendix}: $D=d+\Gamma$ or ($D=d+\omega$) {\it is the local version of} $\nabla_Xs$

\

Let $\xi_V:V-E\buildrel{\pi_V}\over\longrightarrow M$ be a real vector bundle on $M$, $\Gamma(E)$ and $\Gamma(TM)$ the sections of $E$ and of the tangent bundle $TM$ of $M$, respectively, and $C^\infty(M)$ the smooth real valued functions on $M$. A {\it connection} on $\xi_V$ is a function $$\nabla:\Gamma(TM)\times\Gamma(E)
\to\Gamma(E), \ (X,s)\mapsto \nabla (X,s)\equiv\nabla_Xs$$ with the following properties:

\

i) $\nabla_{X+X^\prime}s=\nabla_Xs+\nabla_{X^\prime}s$

\

ii) $\nabla_{fX}s=f\nabla_Xs$

\

iii) $\nabla_X(s+s^\prime)=\nabla_Xs+\nabla_Xs^\prime$

\

iv) $\nabla_X(fs)=X(f)s+f\nabla_Xs$

\

where $f\in C^\infty(M)$. $\nabla_Xs$ is {\it the covariant derivative of} $s$ {\it with respect to} $\nabla$ {\it in the direction} $X$. This is a global definition, and in Physics it describes the interaction between the ``matter'' field $s$ and the ``gauge'' field $\nabla$.

\

Let $m$ and $n$ be the real dimensions of $V$ and $M$, respectively. Let $U$ be an open subset of $M$, $x^\mu$, $\mu=1,\cdots,n$ coordinates on $U$, and $\sigma_i$, $i=1,\cdots,m$ a basis of local sections of $E$. Then, locally, $$\nabla_Xs=\nabla_{\sum_{\mu=1}^nX^\mu\partial_\mu}(\sum_{i=1}^ms^i\sigma_i)=\sum_{\mu=1}^n\sum_{i=1}^mX^\mu\nabla_{\partial_\mu}(s^i\sigma_i)=\sum_{\mu=1}^n\sum_{i=1}^mX^\mu((\partial_\mu s^i)\sigma_i+s^i\nabla_{\partial_\mu}\sigma_i)$$ $$=\sum_{\mu=1}^n\sum_{i=1}^m((\partial_\mu s^i)\sigma_i+s^i\sum_{j=1}^m\Gamma^j_{\mu i}\sigma_j)=\sum_{\mu=1}^n\sum_{i,j=1}^m X^\mu(\delta^j_i\partial_\mu +\Gamma^j_{\mu i})s^i\sigma_j\equiv \sum_{\mu=1}^n\sum_{i,j=1}^mX^\mu D^j_{\mu i}s^i\sigma_j$$ where $$\Gamma^j_{\mu i}=\nabla_{\partial_\mu}\sigma_i$$ are the Christoffel symbols of the connection -its local version- and $$D^j_{\mu i}=\delta^j_i\partial_\mu+\Gamma^j_{\mu i} \eqno{(*)}$$ is the local covariant derivative operator. Multiplying $(*)$ by the 1-forms $dx^\mu$ one obtains the $n\times n$ matrrix of 1-forms $$D^j_i=\delta^j_id+\Gamma^j_i$$ where $d=dx^\mu\partial_\mu$ is the De Rahm operator. The structure $$D=d+\Gamma$$ or $$Ds=(d+\Gamma)s$$ is called {\it minimal coupling} between $\Gamma$ and $s$.

\

{\it Remark}: The more usual definition of connection is as a distribution $H$ of horizontal spaces $\{H_p\}_{p\in P}$ in the total space of a principal $G$-bundle $\xi:G\to P\buildrel{\pi}\over\longrightarrow M$, to which $\xi_V$ is associated through a left action $G\times V\to V$. At each $p\in P$, $H_p\oplus V_p=T_pP$, where $V_p$ is tangent to the fiber $P_{\pi(p)}$, and $\pi_{*p}(H_p)=T_{\pi(p)}M.$ Also, $\psi_{g*}(H_p)=H_{pg}$ where $\psi$ is the right action of $G$ on $P$, with $\psi_g(p)=\psi(p,g)$. There is an equivalence between $\Gamma(E)$ and functions $\gamma:P\to V$ with $\gamma(pg)=g^{-1}\gamma(p)$: $s\mapsto \gamma_s$ and $\gamma\mapsto s_\gamma$. Then, $\nabla_Xs=s_{X^\uparrow(\gamma_s)}$, where $X^\uparrow$ is the horizontal lifting of $X$ by $H$.

\

* On sabbatical leave from UNAM, M\'exico 

\

e-mail: socolovs@nucleares.unam.mx

\end